\documentclass[]{mn2e}

\usepackage{amssymb}
\usepackage{graphicx}
\usepackage{amsmath}
\usepackage{graphics}

\usepackage{color}

\title[Alignment of molecules] {On the dissipation of the rotation energy of dust grains in interstellar magnetic fields}

\author[R. Papoular]{R. Papoular$^{1}$\thanks{E-mail:papoular@wanadoo.fr}\\
$^{1}$Service d'Astrophysique and Service de Chimie Moleculaire,\\
CEA Saclay, 91191 Gif-s-Yvette, France}

\begin{document}

   \maketitle
\label{firstpage}

\begin{abstract}
A new mechanism is described, analyzed and visualized, for the dissipation of suprathermal rotation energy of molecules in magnetic fields, a necessary condition for their alignment. It relies upon the Lorentz force perturbing the motion of every atom of the structure, as each is known to carry its own net electric charge  because of spatial fluctuations in electron density. If the molecule is large enough that the frequency of its lowest-frequency phonon lies near or below the rotation frequency, then the rotation couples with the molecular normal modes and energy flows from the former to the latter. The rate of this exchange is very fast, and the vibrational energy is radiated away in the IR at a still faster rate, which completes the removal of rotation energy. The energy decay rate scales like the field intensity, the initial angular velocity, the number of atoms in the grain and the inverse of the moment of inertia. It does not depend on the susceptibility. Here, the focus is on carbon-rich molecules which are diamagnetic. The same process must occur if the molecule is paramagnetic or bathes in an electric field instead.

A semi-empirical method of chemical modeling was used extensively to illustrate and quantify these concepts as applied to a hydrocarbon molecule. The motion of a rotating molecule in a field was monitored in time so as to reveal the energy transfer and visualize the evolution of its orientation towards the stable configuration. 

\end{abstract}


\section{Introduction}

The observed polarization of light emitted or intercepted by dust in magnetic fields has been attracting attention for the last six decades. A number of interpretations of this phenomenon were offered (see, for instance, reviews and bibliography by Spitzer 1978, Hildebrand 1988, Roberge 2004, Lazarian and Cho 2005 and Lazarian 2007).  In their pioneering work, Davies and Greenstein \cite{dg} (to be referred to below as DG), set the stage for later discussions. Their scenario considers a rigid, solid state grain which, somehow, has been given an angular momentum distributed over its principal axes. As a result, it precesses and nutates about the angular momentum, which  must be conserved absent any other perturbation, as illustrated by the Poinsot construction (Goldstein 1980).  Since the momentum is not, in general, parallel to the ambient magnetic field, $\vec B$, some way must be found for at least those  of its components that are not parallel to $\vec B$ to decrease to zero. In the DG paradigm, the grain is paramagnetic, and the imaginary component of its susceptibility induces a retarding torque. Its braking effect is slower the higher the moment of inertia, so the grain ends up rotating about its highest moment of inertia, the latter being parallel to the field, which is the orientation corresponding to the observed polarization.
 
 This paradigm has been elaborated on (see Jones and Spitzer 1967 and Spitzer 1978, Purcell 1969 and 1979, for instance) and several variants were later proposed, especially in relation to the observed so-called anomalous microwave emission (AME), in particular those involving rapidly spinning ultra small grains (see Draine and Lazarian 1998, Hoang and Lazarian 2014, Ysard and Verstraete 2010,  Silsbee and Draine 2015).
 
 Here, the subject is restricted to the rotation energy dissipation. In this respect, the papers by Purcell \cite{pur79} and Rouan et al. \cite{rou} are remarkable by the breadth of range of subjects discussed. In particular, Purcell suggested imperfect elasticity (internal friction) as a very efficient mechanism: internal heat is produced at the expense of rotation energy, as a result of the periodic changes in internal mechanical stresses due to centrifugal forces as the rotation axis wanders around the grain body. Rouan et al. studied the quantum limit of this mechanism for small molecules, which they call Internal Vibration-Rotation Energy Transfer (IVRET), based on the work of Nathanson and McClelland \cite{nat}. Purcell \cite{pur79} also considered conversion of rotation energy into internal heat through the Barnett effect, which is the inverse of the Einstein-de Haas effect for para- and ferro-magnetic bodies.

Now, iron's cosmic abundance is relatively low and this element is known to associate preferentially with silicon and oxygen. On the other hand, carbon dust is one of the two main components of cosmic dust, often more abundant than oxygen-rich dust, and it is responsible for a much richer IR spectrum. It is therefore necessary to study its alignment too.  In the absence of para- or  ferromagnetism, we are left with diamagnetism as the main remaining plausible cause of alignment. Long ago, Pascal \cite{pas} showed that a diamagnetic susceptibility is associated with the electron pairs that bond atoms in molecules. While this magnetism is much weaker than the paramagnetism associated with the electron spins of unpaired bonds or some heavy atoms, it is more common and stronger in aromatic molecules. Moreover, Pascal's work and  methods are still being discussed and updated with new data (see Bain and Berry, 2008). It is therefore worthwhile exploring diamagnetism, especially in view of the abundance of carbon-rich molecules.

The present work takes advantage of the availability of chemical modeling methods which allow for externally applied electric and magnetic fields (e.g. semi-empirical TNDO or Typed Neglect of Differential Overlap). Using such methods, it becomes possible to explore a large number of structures, rotating or otherwise, measure their diamagnetism, and study their dynamics in a magnetic field.

The main result of this study is a new scheme for the dissipation of dust grain rotation energy, a necessary condition for alignment. This starts with the observation that, when a carbon molecule rotates or oscillates in a magnetic field, its normal mode vibrations are excited to various extents, all the more the stronger the field. This can be understood in several ways. First, such molecules being electrically conducting to some extent, eddy currents are created in them as they move through the field. The heat generated by these currents takes the form of internal vibrations. More basically, different atoms of the structure carry different net local electric charges although the molecule is neutral as a whole. The motion of an atom across the field, induces an electromagnetic force (Lorentz force), $q\vec V \times \vec B$, orthogonal to both and acting upon the atom. As this force varies periodically, the system is reminiscent of two detuned coupled resonant circuits (see Louisell  1960). If the molecule is large enough that its lowest frequency normal mode of vibration is lower than the rotation frequency,  then energy begins to flow from the rotation motion into internal vibrations. Still another way to understand the phenomenon is to look at the periodic impulses given by the field to the atomic charges as random impacts of free flying ambient atoms, whose net result is a friction force retarding the angular motion.

As the internal vibrations grow stronger, the induced molecular magnetic moment acquires an increasing random component, giving rise to a random torque which tilts the angular momentum in random successive directions. This is reminiscent of a particle traveling with a high initial velocity through a fluid whose particles brake the initial linear momentum. In the case of interest, both secular and random components of the angular momentum decrease as the internal vibrations convert into IR radiation.

 The molecule is then attracted into a state where its residual angular momentum becomes parallel to $\vec B$. For the total energy to be a stable minimum, this orientation must correspond to a maximum of induced magnetic moment (maximum susceptibility); this generally coincides with one of the higher inertia principal axes.  The whole process is very similar to the behavior of an axially symmetric top in the terrestrial gravitation field.

The mechanism is governed by the same equations as in (9) and (10) of DG:

\begin{equation}
\dot R=\vec L\cdot\vec\omega\,\,;\,\, \dot\vec{J}=\vec L
\end{equation}

where $R$ is the rotational kinetic energy ($I\omega^{2}/2$ in the simple case of a spherical grain), $\vec J$ the angular momentum, and $\vec L$, the retarding torque, except that, here, the expression of $L$ must be amended. Since, according to the present paradigm, the braking is mediated by the Lorentz force, and the rate of energy transfer scales like the angular frequency, $L$ must include a factor $\omega^{2}B$. This will be developed in Sec. 3.2 and 3.3.

Chemical modeling helps illustrating and quantifying this concept as applied to specific molecules (Sec. 2). In particular, it allows one to define the essential dependence of $L$ upon the various parameters. Unfortunately, present algorithms become very demanding in computation time when a magnetic field is switched on. Fortunately, much less so for electric fields. Since carbon-rich molecules usually carry a finite electric moment (an expression of the non-uniformity of the electronic charge distribution), the dynamics in an electric field is qualitatively similar to that in a magnetic field, differing only by the fact that the force on a given atom is $q\vec E$ instead of  $q\vec V\times\vec B$. Using an electric field allows one to demonstrate the concept as applied to a larger molecule so its lowest vibration frequency comes closer to the rotation frequency (Section 3).

In Sec. 4, we revert to magnetic fields applied to a smaller molecule, first to study the magnetic susceptibility, then to illustrate the dynamics and demonstrate the braking mechanism, as in an electric field.  Finally, as a byproduct of simulations, two types of drifts of the molecule as a whole are illustrated in the Appendix.

\section{The chemical modeling apparatus}

There are several ways in which modeling is essential to the present purposes. First, while data relative to the paramagnetism of solids are quite abundant, this is not the case  for  diamagnetism, if only because it is much weaker and, hence, difficult to measure. When it comes to particular IS molecules, such knowledge barely exists. It can be delivered by chemical modeling with reasonable accuracy. Modeling is also the only way to estimate other properties of the selected molecules, such as the moments of inertia, the net electronic charges and dipole moments, the frequency and energy of rotation-inducing events, etc. which also affect the final degree of molecular alignment. In the present instance, modeling proved to be particularly valuable in avoiding the usual far-reaching approximation of treating the dust particle as a solid non-deformable body with a simple, symmetrical shape. This made it possible to demonstrate the coupling between bulk rotation and internal vibrations.

The principles of chemical modeling in presence of a magnetic field will now be briefly outlined (see Van Vleck 1932, Kittel 1955)
. The dynamics of a molecule is controlled by the electric and magnetic fields, $\vec E$ and $\vec H$, which in turn obey Maxwell's equations

\begin{equation}
\vec E=-\nabla V-\frac{\delta A}{\delta t}   ;   \\
\vec B=\nabla \times \vec A,
\end{equation}

 where $V$ and $\vec A$ are the scalar and vector potentials, and $\vec B$ the magnetic induction. Assuming the magnetic field to be uniform and pointing to the z direction, the components of $\vec A$ are found to be

$ A_{x}=-\frac{1}{2}yH,        A_{y}=\frac{1}{2}xH,         A_{z}=0. $

As a result, the dynamical relations for an atom of the structure, at distance $\vec r$ from the origin, become (in the Gauss system)

$\vec p=\vec p_{kin}+\vec p_{pot}= m\dot\vec r+e\vec A/c $

for the generalized momentum and, for the kinetic energy,

$T=\frac{1}{2}m\dot r^{2}= \frac{1}{2m}p^{2}-\frac{e}{mc}\vec p.\vec A+\frac{e^{2}}{2mc^{2}}A^{2}.$

 Since the quantum mechanical operator for the momentum is  $-\mathrm{i} \hbar\nabla $,  the corresponding perturbation potential to be added to the Hamiltonian $H$ of the atom (not to be confused with the magnetic field) is 
 
 $\it H =\frac{\mathrm{i}e\hbar}{2mc}(x\frac{\delta }{\delta  y}-y\frac{\delta }{\delta x})+\frac{e^{2}H^{2}}{8mc^{2}}(x^{2}+y^{2}).$
 
 Several commercially available chemical modeling packages provide for dynamics computations without magnetic field, in which case $\vec A$=0 and $V$ reduces to the instantaneous electrostatic potential. The Hyperchem package used here (available from Hypercube, Inc., USA)  offers the magnetic option only for some theoretical methods which neglect some of the interactions between electrons of the structure (NDO semi-empirical methods:  Neglect of Differential Overlap).  Here, we use the TNDO (Typed NDO) method, which is considered as the currently most elaborate semi-empirical method. The general capabilities of the package (drawing, energy optimization, computation algorithms, molecular properties, molecular dynamics, atomic impacts, etc.) have been described elsewhere (see Papoular 2005, and references therein). 
 
\begin{figure}
\resizebox{\hsize}{!}{\includegraphics{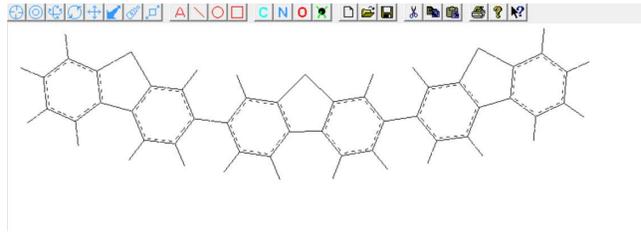}}
\caption[]{The carbon-rich molecule used in this work to illustrate the argument: 59 atoms, 36 C, 20 H, 3 S (at the pentagons free summits). Its 3 principal axes coincide with axes Ox, Oy, Oz of the figure. The electric dipole moment runs through the center of mass, pointing downwards because of the excess electronic charges on the sulfur atoms.}
\label{Fig:ttri}
\end{figure}

 While IS dust is made of an infinite variety of molecules with a wide distribution of sizes, most of the properties and interactions we wish to demonstrate here can be illustrated using the same, medium-size, representative ``generic'' molecule which does not require excessive computational times. This is sketched in Fig. \ref{Fig:ttri} and will sometimes be referred to as the ``3trioS''. A simple organic molecule was chosen, as such molecules have long been known for their strong diamagnetism (P. Pascal 1910, Pauling 1936). Any of the following modeling experiment can be repeated on any molecule of reasonable size.

 Some characteristics of the 3trioS are as follows:

-Mass 548 a.u.; Binding (potential) energy: -7219.04 kcal/mole (1 kcal/mole=1/23 eV per molecule); Electric Dipole Moment: 3.76 Debye  (1 Deb=$10^{-18}$ escgs).

-Minimum normal mode wavenumber: 6.64 cm$^{-1}$ ($2\,10^{11}$ Hz).

-Mean polarizability under electric field of $5\,10^{-4}$ a.u. ($2.57\,10^{5}$ V/cm):  65.4 $\AA{\ }^{3}$.

-Partial atomic (Mulliken) charge, an expression of the local fluctuations of the net electric charge and potential energy: excess electronic charge from +0.245 e on the sulfur atoms, to -0.025 e on carbon atoms.

-Principal moments of inertia along Ox, Oy, Oz (axes 1, 2, 3): 1188.09, 27622.8, 28810.9  $\times\,\,1.7\,10^{-40}$ g.cm$^{2}$ .

 The 3 principal axes of inertia are designated by 1, 2 and 3 (set here along the figure axes Ox, Oy and Oz, respectively), and run through the CM (center of mass). The DM (electric dipole moment) also runs through the CM, pointing down. Its finiteness indicates that the centers of positive and negative net atomic charges do not coincide; its value is the same as that of a dipole of length 1 $\AA{\ }$ and charge 0.65 electronic charge. This plays an essential role in molecular dynamics under electric or magnetic fields.

\section{Dynamics in an electric field}

\subsection{Some characteristic times}
As several processes are operating simultaneously in the IS medium (ISM), their characteristic times must be compared in order to properly set the stage for the simulation experiments. Assume the flux of dissociating UV photons to be $10^{7}$ photons.cm$^{-2}$.s$^{-1}$, typical of the tenuous, high-latitude ISM (G=1; see Draine 1978) . Take the UV absorption cross-section of carbon atoms to be $10^{-17}$ cm$^{2}$, and the average number of such atoms per dust particle to be 1000. Then, the  average interval between UV impacts is about a year. Likewise, the average interval between H atom impacts is about 10$^{6}$ s if the ambient density of such atoms is 20 cm$^{-3}$, their speed $10^{5}$ cm.s$^{-1}$ and the effective cross-section of the target, 100$\times$100 $\AA{\ }^{2}$. Since the relaxation time for the grain to settle into a state of statistic vibrational equilibrium does not exceed 1 $\mu$s, and the IR radiative lifetime is less than 1 s, it is clear that, most of the time, the grain's internal temperature is zero, although, on average, it may reach 20 K.

However, upon an H atom impact, the probability that a H$_{2}$ molecule is formed and ultimately ejected is not negligible (e.g. see Papoular 2005). The attendant rocket effect then sends the grain into  supra-thermal rotational motion. Our problem then simplifies into studying what happens to such a rotating grain in a magnetic field, in the time interval between such rocket effects.

An analysis of the energy budget of this process (\it op cit. \rm) shows that most of the released energy goes into the kinetic energy of linear, rotational and vibrational motions of the hydrogen molecule. The energy left in the dust grain is usually less than 1 kcal/mole (0.05 eV).  The resulting angular velocity, $\omega$, is generally taken to be  as high as 10$^{9}$ s$^{-1}$ (see Spitzer 1978 ) for grain sizes of $10^{-5}$ cm. In the simulation experiments below, with smaller molecules, this will be set in the order of $10^{12}$ s$^{-1}$, so as to keep the computation time within reasonable limits.

\subsection{The braking process}
As mentioned above, the software cannot handle sufficiently large molecules in a magnetic field. This is not the case with electric fields. Since the proposed braking process relies on the excitation of internal vibrations by the force exerted by the field on individual atomic charges, it does not make a fundamental difference whether this force is $q\vec E$ or $q\vec V\times\vec B$.  Accordingly, the proposed braking process will first be demonstrated on a molecule in an electric field ( absent any external magnetic field.)

The molecules we are interested in usually have a significant electric dipole moment. As soon as the field is turned on, the molecule generally starts rotating in response to the couple formed by the field and the induced or permanent dipole moment. One can either: a) start the experiment with the molecule at rest, so it is first accelerated then braked by the field, or b) first let the molecule reach the desired angular velocity in a predetermined molecule/field configuration, then set the field to the desired value and start monitoring the braking. Both yield similar results. 

The action of rotation upon the atoms is best evidenced by starting with the molecule at rest in the plane of the figure (Fig. \ref{Fig:ttri}) and the field along Oz, orthogonal to it; the torque on the dipole moment is then at its maximum and the molecule begins to rotate as soon as the field is turned on. Angular rotation and momentum are both parallel to Ox. The field intensity is set at 0.01 a.u. (1 atomic unit=5.2 $\,10^{11}$ V/m). This is a huge field by ISM standards, but it is unavoidable if the effect is to be demonstrated with available modeling software and reasonable computation times. Repeating the experiment with fields between  0.01 and 0.001 a.u. suggests that the braking time  is inversely proportional to the field strength; this makes sense if the coupling between rotation and internal vibrations is due to the electrostatic force acting on the atoms. This scaling rule will be used below for extrapolations to the ISM.

\begin{figure}
\resizebox{\hsize}{!}{\includegraphics{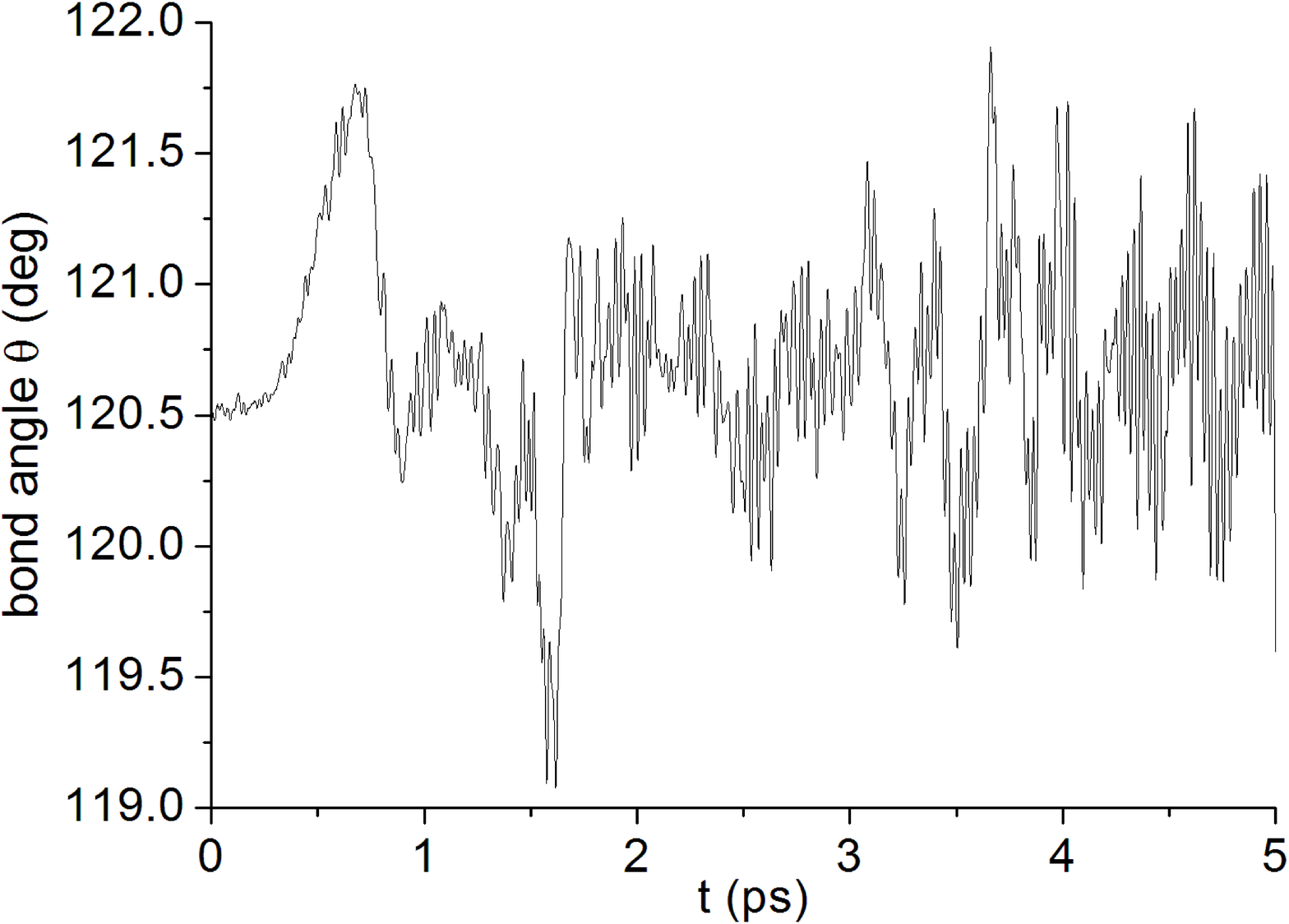}}
\caption[]{Time evolution of the angle between a CH bond and an adjacent CC bond, from the moment the field is turned on and the molecule starts rotating about Ox. Note the change in character of the motion, from the response of the H atom alone to the initial electric kick, during the first picosecond, to the later, much faster oscillations including in-plane and out-of-plane CH bending modes.}
\label{Fig:startB}
\end{figure}

\begin{figure}
\resizebox{\hsize}{!}{\includegraphics{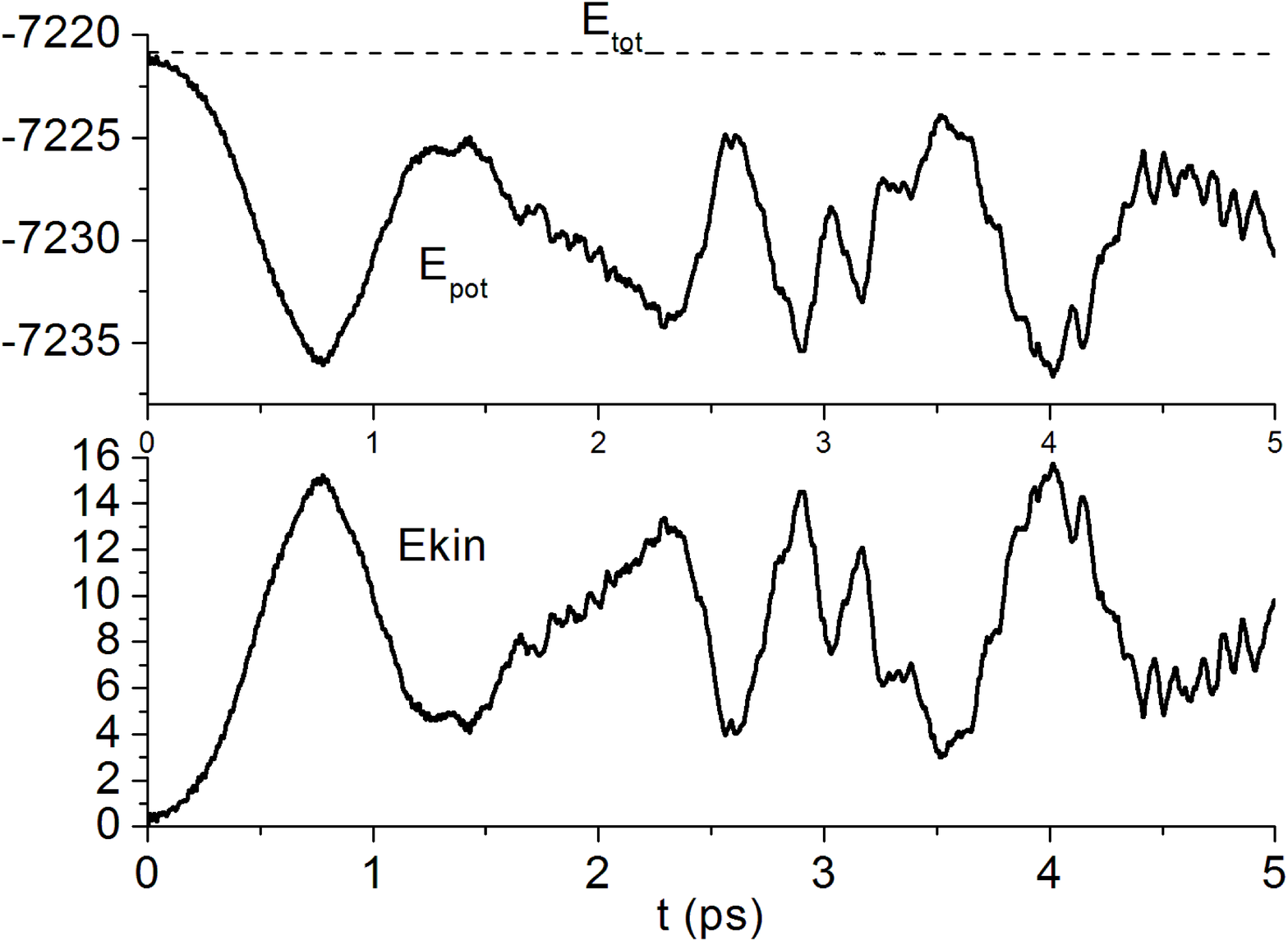}}
\caption[]{Total molecular bonding (Epot) and kinetic (Ekin) energies, and their sum (Etot), in units of kcal/mole as a function of time, starting with application of the field. The variations of the two quantities are exactly out of phase, as expected from energy exchanges between different molecular degrees of liberty, excluding a field contribution.}
\label{Fig:startEpEk}
\end{figure}

The software provides several means by which to monitor the dynamics of the molecule and make the braking apparent. One of the most telling quantities that can be monitored is the angle between a CH bond and an adjacent CC bond. For the nearest CH bond to the right of the central S atom, this angle is $\theta$=120.5 ° at rest in the optimized configuration with no field. Its evolution starting from the moment when the field is turned on is illustrated in Fig. \ref{Fig:startB}. The starting kick has given the H atom a strong linear momentum, which is converted, within 1 ps,  into normal mode vibrations, mainly the o.o.p. (out-of-plane) bending modes, in the 13-$\mu$m band and i.p (in-plane) bending modes in the 7-$\mu$m range. \it This is the core of the proposed braking mechanism. \rm

Other significant markers are the total kinetic energy, $E_{kin}$, and total bond energy, $E_{pot}$, (Fig. \ref{Fig:startEpEk}). Both include the rotational and vibrational energies. Initially, the rotational energy is dominant, so its variations in time roughly follow the variations of the electrostatic torque with the angular position, and the ensuing maximums of potential energy, coinciding with minimums of kinetic energy and \it vice versa\rm, at intervals determined by the acquired angular velocity. Thus, the first peak of kinetic energy, $\sim15$ kcal/mole at $\sim0.8$ ps, coincides in time and amplitude with the first dip in potential energy. At that point, the molecule has acquired its maximum rotational speed,  $\sim3.3\,10^{12}$ rad/s (17.5 cm$^{-1}$).

\begin{figure}
\resizebox{\hsize}{!}{\includegraphics{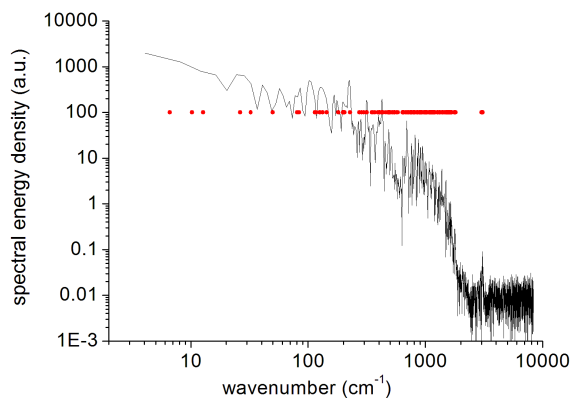}}
\caption[]{Black line: Fourier transform of the kinetic energy variations from 5 to 10 ps after staring the experiment. Nearly all normal modes (red dots at ordinate 100) are excited, with the energy flowing from low to high frequencies, as expected if the source is the rotation at a frequency of 17.5 cm$^{-1}$.}
\label{Fig:specEkin}
\end{figure}

After a few such cycles, however, the amplitude of these energy variations decreases visibly and gives way to weaker, faster and less regular variations as more and more rotational energy is converted into internal vibrations. This is confirmed by monitoring the molecule on the screen all the while: one can visualize the undulations of the structure characteristic of those of the normal modes which have the lowest frequencies (phonons), as described in Papoular \cite{pap15}.

Note that, all along, the total energy (sum of the kinetic and bond energies) remains essentially constant, confirming that the energy exchanges are exclusively internal and not with the electric field. The only contribution of the latter is in the form of the potential energy which was deposited the moment the field was turned on.

Details of the energy transfer can be quantified by taking the Fourier transform (FT) of either $E_{kin}$ or $E_{pot}$. Figure \ref{Fig:specEkin} displays the FT of the kinetic energy variations from 5 to 10 ps after launching the experiment. It all but coincides with the FT of the bond energy variations, except beyond 2000 cm$^{-1}$, where they differ only in background continuum. Comparison with the normal mode spectrum of the molecule (red dots at ordinate 100) indicates that most modes have been excited. The general slope of the spectrum suggests that energy has been flowing from low to high frequencies, as expected if the source is the rotation energy variations at a frequency of 17.5 cm$^{-1}$.

\begin{figure}
\resizebox{\hsize}{!}{\includegraphics{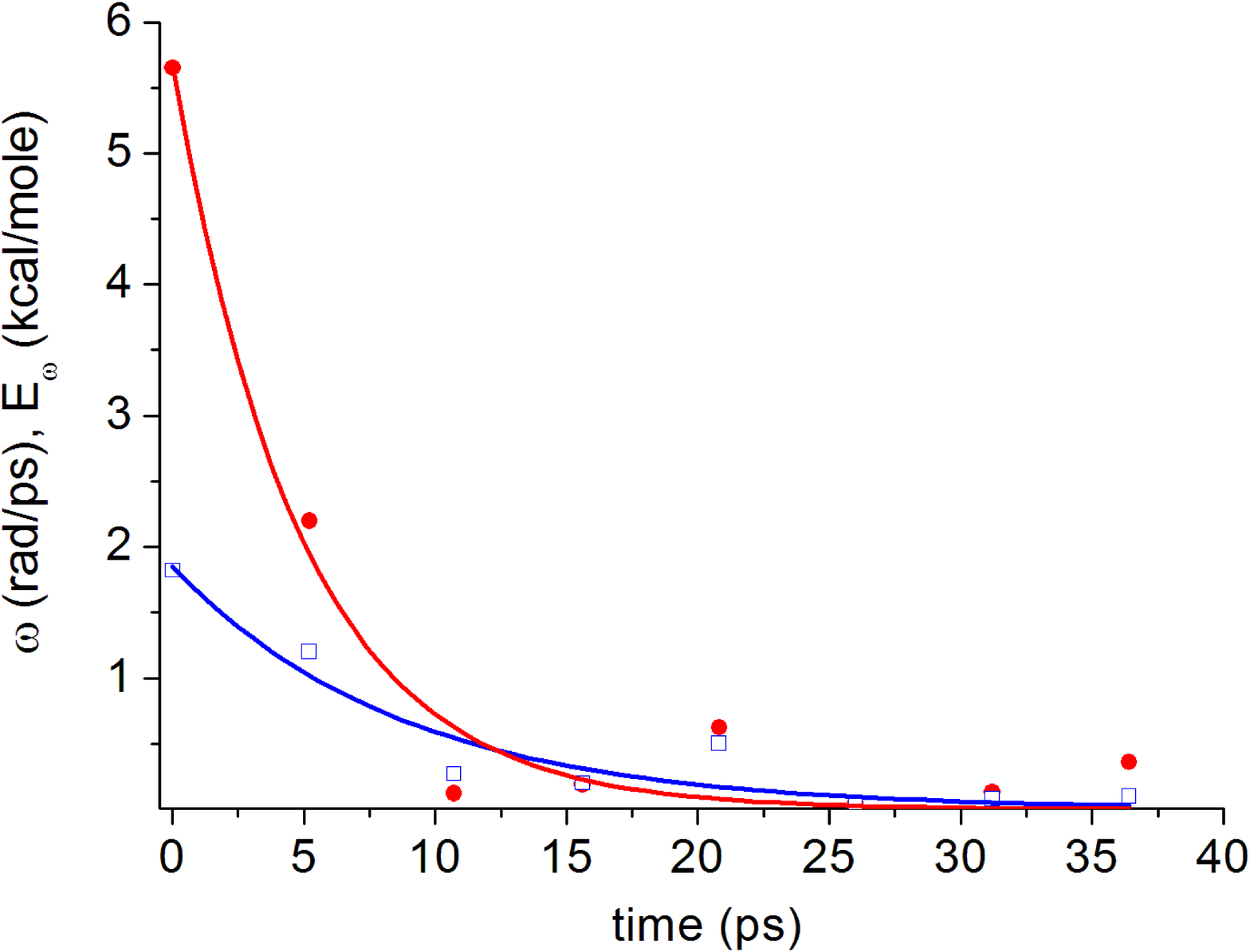}}
\caption[]{Angular frequency $\omega$ (open blue squares) and rotational energy (red dots) as a function of time. The time constants of the superposed best fitting exponential functions are, respectively, 8.7 and 4.9 ps. Here, the molecule was initially given an angular frequency 1.82 rad/ps parallel to its principal inertial axis 1 and to coordinate axis Ox, with a rotational energy 5.65 kcal/mole. The electric field was then set at 0.003 a.u. parallel to Oz.}
\label{Fig:OmegaE}
\end{figure}

That the vibrational energy increase is at the expense of the angular velocity and rotational energy is illustrated by another experiment. Here, the molecule was initially given an angular frequency 1.82 rad/ps parallel to its principal inertial axis 1 and to coordinate axis Ox, with a rotational energy 5.65 kcal/mole. The electric field was then set at 0.003 a.u. parallel to Oz. Fig. \ref{Fig:OmegaE}  displays those two quantities as a function of time during the following 37 ps. The time constants of the superposed best fitting exponential functions are, respectively, 8.7 and 4.9 ps. The former is here referred to as the braking time (or retarding, or decay time). The dispersion of the data points about the fits is due to the energy transfer being an irregular process which depends not only on the angular velocity, but also on the varying combination of the instantaneous positions and velocities of the vibrating atoms.

\begin{figure}
\resizebox{\hsize}{!}{\includegraphics{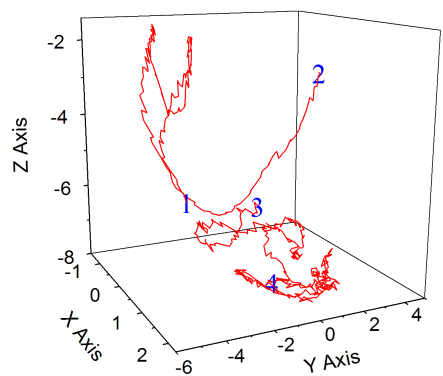}}
\caption[]{Trajectory of the positive end of the electric dipole moment in 3D. Here, the molecule was initially given an angular frequency 1.82 rad/ps parallel to its principal inertial axis 1 and to coordinate axis Ox, with a rotational energy 5.65 kcal/mole. The electric field is then set at 0.003 a.u. parallel to Oz. For clarity, only an initial leg,1-2 (0 to 5.22 ps) and a final leg, 3-4  (20 to 30.57 ps) of the trajectory are shown. The motion is initially pendulum-like about Ox in the yz plane; as rotation energy is converted into internal vibration energy, it becomes more and more irregular, but with a clear trend towards the bottom of the stable potential well created by the field (with its axis of maximum moment of inertia parallel to the field). However, because the final radiative removal of its energy cannot be simulated here, it ends up as pendulum oscillations about Oz, in the xy plane}
\label{Fig:DME}
\end{figure}

This is made clearer in Fig. \ref{Fig:DME}, which displays the trajectory, in 3D, of the positive tip of the electric dipole moment.
 Here again, the molecule is initially given a rotational velocity $\omega=1.8\, 10^{12}$ rad/s, corresponding to a rotational energy of 5.65 kcal/mole (0.25 eV), parallel to Ox, which initially coincides with inertial axis 1.  The electric field is then set at 0.003 a.u. (1.56$\,10^{7}$ V/cm) parallel to Oz, and the electric dipole moment is monitored for the following 30 ps. For clarity, only an initial leg (0 to 5.22 ps) and a final leg (20 to 30.57 ps) are shown. The former shows the initial pendulum oscillations about Ox, in the yz plane. In the later phase, much of the rotation energy has gone into vibrations and the dipole moment tends towards the configuration where energy is minimum. This is the case when the dipole moment (like the axis of symmetry) is parallel to Ox, which is the bottom of the stable potential well of this structure.

However, it is clear that the dipole moment is not at rest in this configuration but oscillates about Oz, in a plane nearly normal to that axis. This is because the vibration energy has grown to a considerable level. In space, this energy would have been radiated away as soon as it was gained. Unfortunately, the available software does not allow for a simulation of this removal, so all the energy converted from rotation to vibration is available for pendulum oscillations, at least for a very long time. 

The demonstration of the braking process is completed by checking that, if the field is null or parallel to both the momentum and rotation vectors,
the initial angular momentum is unaffected, and the rotation goes on indefinitely.

 Returning to eq. (1) with our new understanding of the braking mechanism, we think of a larger grain made of the same structure as in the example above, and express the reasonable assumption that the rate of rotation energy transfer, $\dot R$, is proportional to the electric force, to the angular frequency and to the grain volume (as the field permeates the hole grain). Hence,

\begin{equation}
L=\lambda EN\omega,
\end{equation}
where $N$ is the number of atoms in the molecule: for the sake of clarity, we make no difference between atoms; $\lambda$ accounts for the distribution of charge among atoms and of the structural details, which are assumed to be independent of size. The solution of eq. (1) then becomes
\begin{equation}
\omega=\omega_{0}\mathrm{exp}(-t/\tau)\,\,;\,\,R=\frac{1}{2}I\omega_{0}^{2}\mathrm{exp}(-2t/\tau)\,\,;\,\,\tau=\frac{I}{\lambda EN}.
\end{equation}
In the experiment reported above, $\tau=8.7\,10^{-12}$ s, $E=0.003$ a.u., or $5.1\,10^{4}$ statV/cm, or $1.5\,10^{7}$ V/cm, $N=59$ atoms and $I=1.12\,10^{3}$ a.u., or $2\,10^{-37}$ g.cm$^{2}$. Hence, $\lambda\sim8.5\,10^{-33}$. It is of interest to note that $\tau$ does not depend explicitly on the angular velocity nor on the susceptibility.

\subsection{The braking time in interstellar space} 
In spite of the obvious difference of force directions,there is no essential difference between acting on the atoms of the molecule through the electric or the magnetic force; we also assume that a magnetic field acting alone, is as effective as an electric field acting alone, provided $\mid\Sigma q\vec E\mid=\mid\Sigma q\vec V\times\vec B\mid$. Consider a component of angular velocity orthogonal to the field (as that which is parallel will not be affected). For simplicity, assume all net atomic charges to be equal, and reduce this relation to the corresponding scalar quantities: $\Sigma$E=B$\Sigma V$. Also consider $V$ as some average over a rotation cycle. Then $V_{i}$=r$_{i}\omega$ for atom $i$, and the sum over the grain is $N\omega r_{av}$, where $N$ is the total number of atoms and $r_{av}$, their average distance from the rotation axis.  Finally, define

\begin{equation}
\tau_{0}=\frac{I}{\lambda NBr_{av}\omega_{0}}.
\end{equation}

Then, from eq.(1),
\begin{equation}
\dot\omega=\frac{\omega^{2}}{\omega_{0}\tau_{0}},
\end{equation}

whose solution is

\begin{equation}
\omega=\frac{\omega_{0}}{1+t/\tau_{0}}.
\end{equation}

Here, therefore, the decay is initially exponential with a characteristic time $\tau_{0}$, then progressively turns hyperbolic. \it Note the dependence of $\tau_{0}$ on $B^{-1}$, as opposed to $B^{-2}$ for paramagnetic braking.\rm

That $\tau_{0}$ scales like $I$ does not mean that the molecule will necessarily settle in the configuration where the principal axis of highest inertia is parallel to the field. Rather, it will settle in the most stable magnetic configuration: axis of symmetry parallel to the field. For an axially symmetric ellipsoid, the two configurations coincide.

 If the frequency of successive rocket events, $\nu$, is not much less than $1/\tau_{0}$, then, the variation of $\omega$ with time has no analytical expression; however, it is fairly well described as follows. As recalled in Sec. 3.1, previous experiments show that the rotation energy deposited in the grain by a rocket effect does not exceed 1 kcal/mole ($\sim0.05$ eV) whatever the grain size. To remain on the safe side, take this as the common value, $R_{0}$. If the relevant moment of inertia is $I$, then $R_{0}=\frac{1}{2}I\omega_{0}^2$ and a term $\nu R_{0}$ must be added to the right-hand side of the first eq. (1). The rotational kinetic energy will stop decreasing ($\dot R\sim0$) roughly around the time when this term is equal to $\frac{1}{2}I\omega^{2}$, where $\omega$ is given by eq.(7). This time is $\tau=\tau_{0}/(\nu\tau_{0})^{1/3}$. Thereafter, the angular velocity does not fall below $\sim\omega_{0}(\nu\tau_{0})^{1/3}$.

We have still to specify the scaling law for the grain size. This is done by writing $I\propto N(r^{2})_{av}$ and $N\propto (\bar r)^{3}$ , so, from (5),

\begin{equation}
\tau_{0}\propto\frac{( r^{2})_{av}^{3/2}}{(\bar r)^{3}}\frac{\bar r}{r_{av}}\bar r^{3.5}.
\end{equation}
To a form factor, this scales approximately like the number of atoms, $N$.

 We are finally in a position to deduce, from the above, specific numerical values of $\tau$ for specific cases of interest to astronomy. Take, for instance, the magnetic field intensity in the ISM to be $5\,10^{-6}$ G, and the molecular angular velocity  2 rad/ps, so a typical atom velocity is 400 m/s (for our generic molecule, 3trioS); then the equivalent electric field intensity  is roughly $2\,10^{-7}$ V/m, which is about $10^{16}$ times weaker than the 0.003 a.u. electric intensity used for Fig. \ref{Fig:OmegaE}. If the extrapolation is valid, the braking time in the ISM, for the molecule considered above (59 atoms), would be in the order of $10^{5}$ s.
 
 Depending on the rate of formation of hydrogen molecules on the grain surface, a more or less significant fraction of the larger grains will be excluded from the alignment process. However, the very steep distribution of grains in linear size should greatly mitigate this effect. But this must be left for further scrutiny.
 
\section{Simulating a molecule in a magnetic field}

Repeating, in a magnetic field, the experiments demonstrating the braking and alignment mechanisms as in electric fields would be valuable. However,
as mentioned in the Introduction, using the magnetic field algorithm with the 3trioS molecule requires impossibly long computational times. We are therefore compelled to turn to a smaller molecule: the structure will be reduced to a single one of the 3 components of 3trioS, the central one, called trioS. Organic molecules are generally diamagnetic to some extent, as will now be illustrated in this particular case. Some characteristics of the trioS are as follows:

-Mass 184 a.u.; Binding energy: -2476.5 kcal/mole; Electric Dipole Moment: 2.72 Debye  (1 Deb=$10^{-18}$ escgs).

-Net atomic charges in the same range as 3trioS above.

-Principal moments of inertia along Ox, Oy, Oz (axes 1, 2 3): 318.5, 866.6, 1185.1  $\times\,\,1.7\,10^{-40}$ erg.

\subsection{Diamagnetic susceptibility}

\begin{figure}
\resizebox{\hsize}{!}{\includegraphics{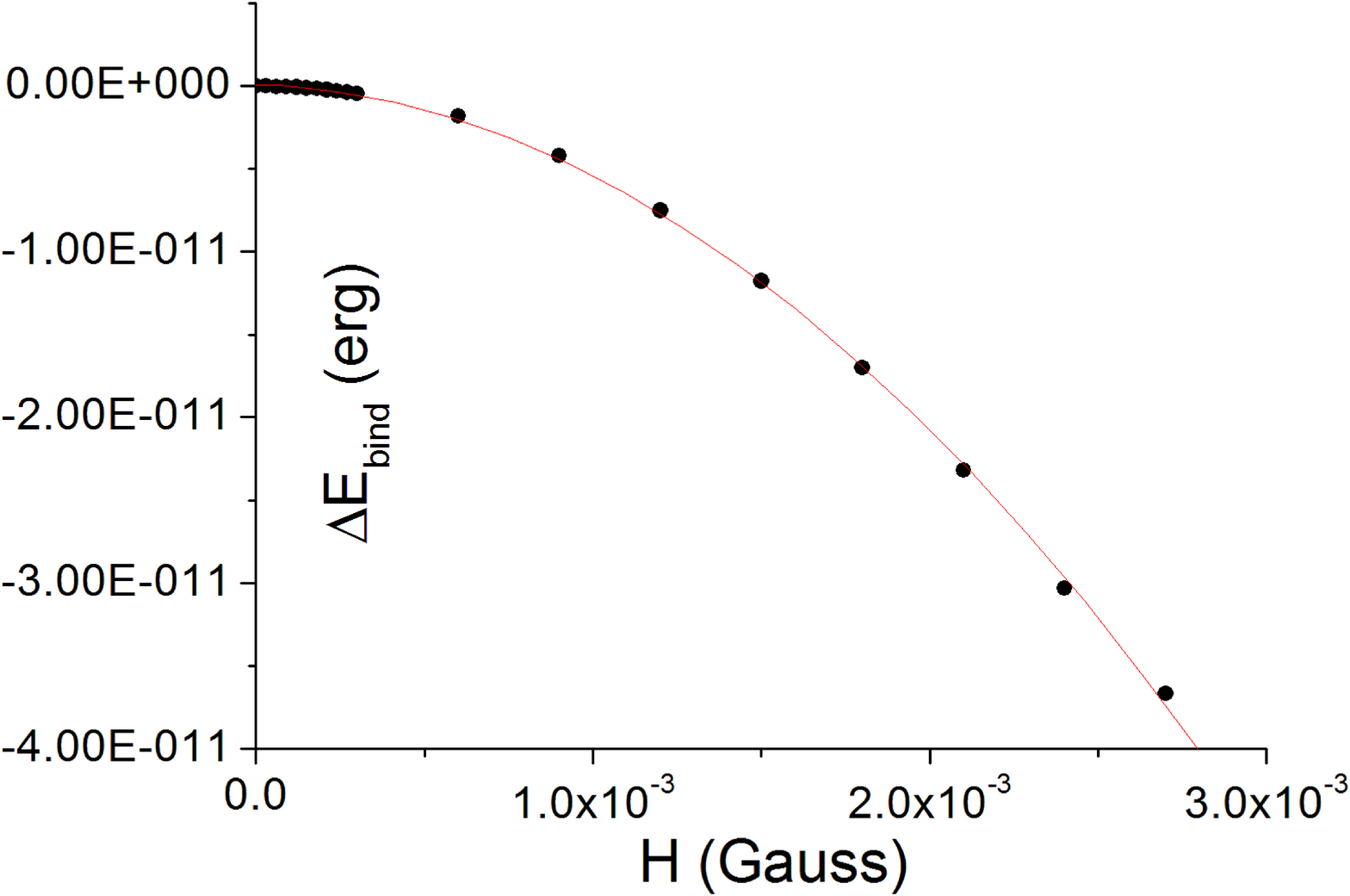}}
\caption[]{Bonding energy increase as a function of applied magnetic field, $H$. The field is applied parallel to the symmetry axis of the molecule (axis 2).}
\label{Fig:magneny}
\end{figure}

 \begin{figure}
\resizebox{\hsize}{!}{\includegraphics{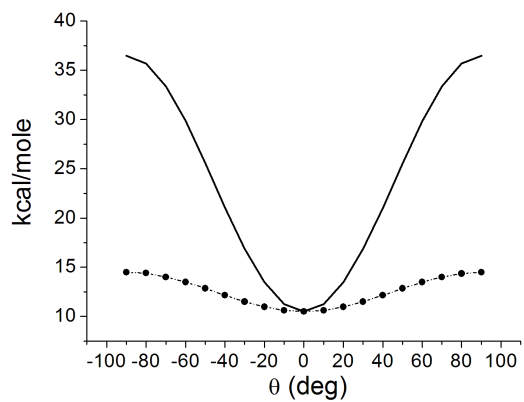}}
\caption[]{Profiles of the magnetic potential well along two orthogonal planes: a) (full) the plane of principal axes 1 and 2 (xy plane); b) (dashed) the plane of axes 1 and 3 (xz plane). $\theta$ is the angle between the symmetry axis and the particular field orientation.}
\label{Fig:potwells}
\end{figure}

 A given material, subjected to a magnetic field, $\vec H$ will develop a magnetic moment, $\vec M$ \it per unit volume \rm, such that
 
 $\vec M=\chi H$, with $\chi=\chi'+i\chi''$.
 
 where $\chi$ is non-dimensional, and $H$ and $M$ are extensive quantities. As $\chi''$ is small, we shall only be interested in $\chi'$ and refer to it as $\chi$.
Now, consider a finite volume, $V$, of this material, possibly a microscopic volume. It will be assumed, for convenience, that the \it total \rm magnetic moment, $M$ (note the change of meaning), of this chunk of material is $\chi HV$. It must be emphasized that $\chi$ is sometimes also given in the chemical physics literature in units of g$^{-1}$,  or mole$^{-1}$ or by only specifying the density of the material (e.g. mole/cm$^{3}$). In each case, the expression for $M$ should be modified correspondingly. Still another case must be confronted when modeling molecules. In that case, $\chi$ must be deduced from the change in bonding (potential) energy upon turning the magnetic field on; this is given by $\Delta E=-\chi H^{2}V$, where $V=1/6\,10^{23}$ cm$^{3}$. The  modeling code gives the field in units of $1.7\,10^{7}$ G (a.u.: atomic units) and the magnetic energy  in kcal/mole (1 kcal/mole is equivalent to 1/23 eV, or $7\,10^{-14}$ erg, \it per molecule).\rm

Thus, the static susceptibility in a given direction can be obtained from the increase in binding energy (algebraic decrease of the molecular potential energy) as an increasing magnetic field is applied parallel to this direction.  This is illustrated in Fig. \ref{Fig:magneny} for direction Oy, which is also the axis of symmetry. A good fit to the points is given by $10^{-5}\,H^{2}V$, so $\chi=10^{-5}$. By comparison, the susceptibilities of the highly diamagnetic materials, graphite and diamond, are of order 2$\,10^{-5}$ (see Young 1992).

The same trend was found for the 2 other principal axes albeit with somewhat different susceptibilities. These 3 particular directions of $\vec H$ define three different local extrema of the potential energy, i.e. three equilibrium configurations. The only stable one occurs when the magnetic field lies along the (in-plane) axis of symmetry of the molecule, perpendicular to its long dimension, as in Fig. \ref{Fig:ttri}. Figure \ref{Fig:potwells} outlines the profiles of the potential well along two orthogonal directions relative to the stable one. Obviously the anisotropy of the well espouses the asymmetry of the molecule. The susceptibility matrix is approximately diagonal and symmetric with respect to the center of mass.

\subsection{Molecular dynamics in a magnetic field}
 In the absence of braking torques, the initial angular momentum imparted to the molecule by previous collision and rocket processes must be conserved, so the molecule can only and indefinitely experience changes in the distribution of the rotation components but never align with the field. This is illustrated by the following experiment.

\begin{figure}
\resizebox{\hsize}{!}{\includegraphics{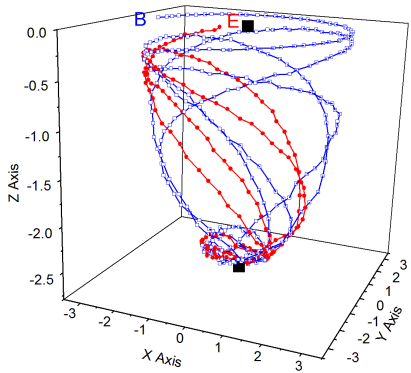}}
\caption[]{Motion of the trioS, initially spinning about its principal axis 3, in a magnetic field pointing in direction Oz, with intensity 20 a.u., as symbolized by the trajectory of the tip of its electric dipole moment. Open blue squares: from 0 (point B) to 10 ps; filled red circles: from 10 to 15 ps (point E). Upper black square: center of the circular initial trajectory. Lower black square: stable target position in this field, for a molecule spinning now about its principal axis 2, parallel to the field. Although the rotation vector continuously wanders over the molecule, from one principal axis to the other, there is no apparent rotation energy dissipation and the dipole moment tip indefinitely pursues the exploration of the same axially symmetric surface.}
\label{Fig:oscil}
\end{figure}
 
To begin with, the molecule must be set in rotation as in the ISM. For this purpose, in the absence of a magnetic field, with the molecule lying in the xy plane symmetrically about Oy, we apply an electric field $\vec E=0.03$ a.u (1 a.u.=$10^{7}$ V/cm) parallel to Ox, i.e. orthogonal to the molecular dipole moment. The resulting torque launches the rotation about axis Oz (parallel to principal molecular axis 3; moment of inertia: 1185 a.u.). After 1/4 of a revolution, the angular velocity, $\omega$, reaches a maximum of 3.36 rad/ps, with an energy of 13.36 kcal/mol. This rotation frequency is much higher than expected in the ISM, but is necessary to enhance all mechanical effects and keep the computation times within reasonable limits.

 The electric field is then turned off and the magnetic field turned on, with an intensity of 20 a.u., parallel to Oz (again, this is much higher than in the ISM, but is also necessary for the same reasons). This configuration is not a stable one, as shown in the previous section; while spinning, the molecule is therefore attracted towards the  stable configuration where its symmetry axis (axis 2; moment of inertia:866.6 a.u.) is parallel (or anti-parallel, because of its diamagnetic character)  to the magnetic field. The motion is governed by the couple between the field and the induced magnetic moment, as well as by the resultant of the Lorentz forces  $q\vec V\times\vec B$ applied to the moving individual electric charges. This is illustrated in Fig. \ref{Fig:oscil}, which displays the trajectory in space of the positive tip of the dipole moment (assumed permanently coincident with axis 2 of the molecule). It begins at B and ends at E: open blue squares from 0 to 10 ps, filled red circles from 10 to 15 ps. The intervals between symbols correspond to 0.03 ps. Initially, the tip describes a circle of radius $\sim2  \AA{\ }$ about the origin of coordinates (upper black square). As it descends towards one of its two stable positions in this field (0, 0, -2.7: lower black square), the radius of the circles decreases. However, it only reaches an altitude of -2.6, and before it can go further, it is attracted back upwards with increasing radius, only to return later, always riding on the same 3D surface. At constant total energy,  a descent towards lower potential energies is accompanied by an increase in kinetic energy which implies an increase in rotation velocity and a decrease of the moment of inertia. Figure \ref{Fig:oscil} shows just this. The final alignment can only be reached if a braking mechanism enters the picture, to absorb enough of the kinetic energy so the molecule can settle in the most stable magnetic configuration. Here, apparently, the anelestacity invoked by Purcell \cite{pur79} is not sufficient to do the job, in spite of the very high intensity of the field. Nonetheless, observation of the molecule on the screen reveals indeed its deformations (phonon modes) under the periodic centrifugal stresses it experiences.

\begin{figure}
\resizebox{\hsize}{!}{\includegraphics{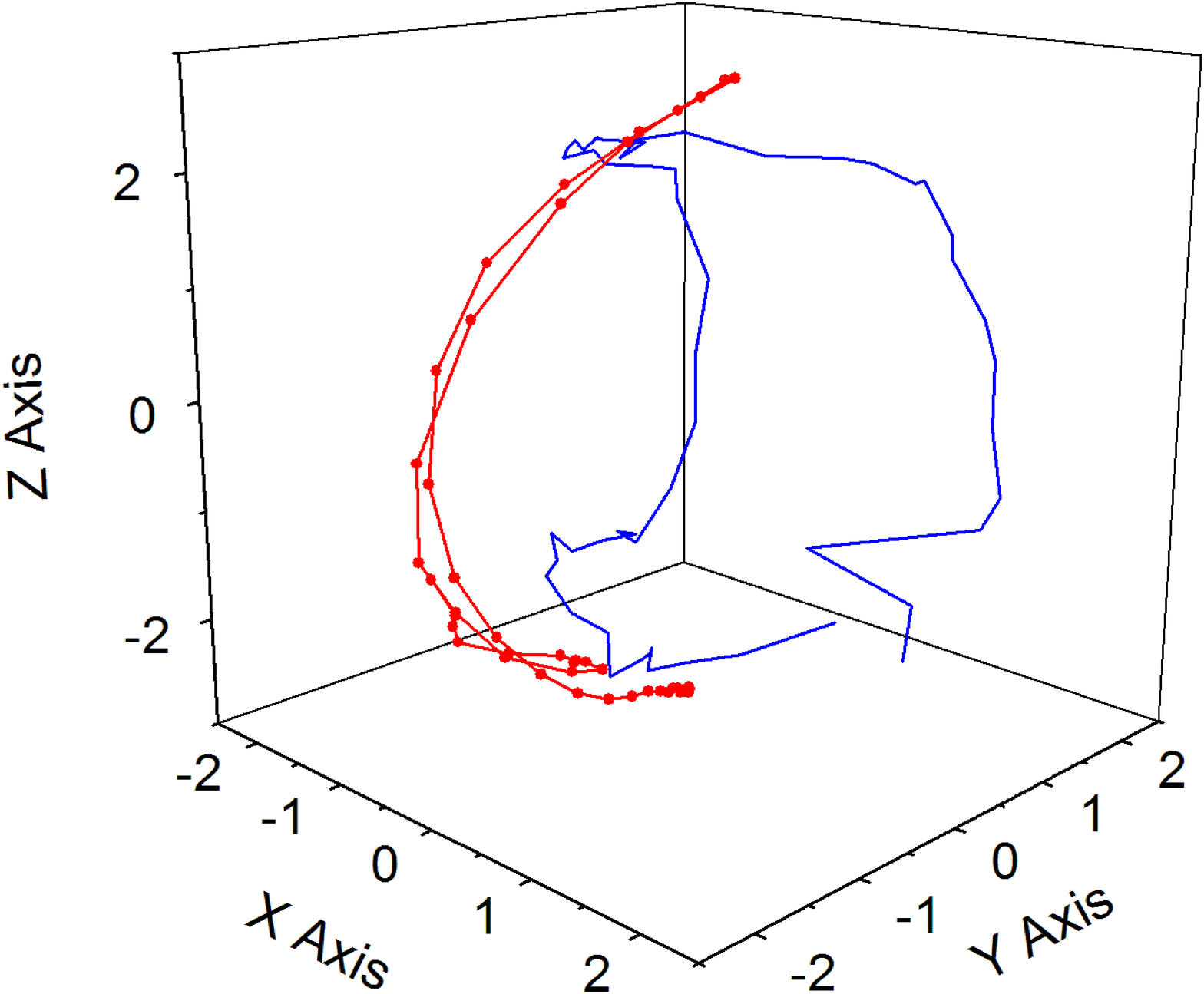}}
\caption[]{Trajectory of the positive tip of the dipole moment of a trioS in a magnetic field of 50 a.u. along Oy. Red curve with dots: from 0 to 1.5 ps (first pendulum oscillation about Ox); blue line: 13.5 to 15 ps (final random motion around the bottom of the stable potential well). See Sec. 2 for comments.}
\label{Fig:DMC12}
\end{figure}

As suggested above, the braking mechanism invoked here relies on the mediation of the Lorentz force $q\vec V\times\vec B$, acting on the individual net atomic charges in the same way as the $q\vec E$ force did above. As the molecule rotates, each atomic charge feels a field varying at the rotation frequency, which periodically perturbs its rotational motion. If this frequency falls within, or not far from, the  internal vibration spectrum of the molecule, then the latter is excited and energy can flow from rotation to vibrations and soon be radiated away. With the available simulation software, we can illustrate the conversion of rotation to vibration energy (but not its evacuation), which we proceed to do.

\begin{figure}
\resizebox{\hsize}{!}{\includegraphics{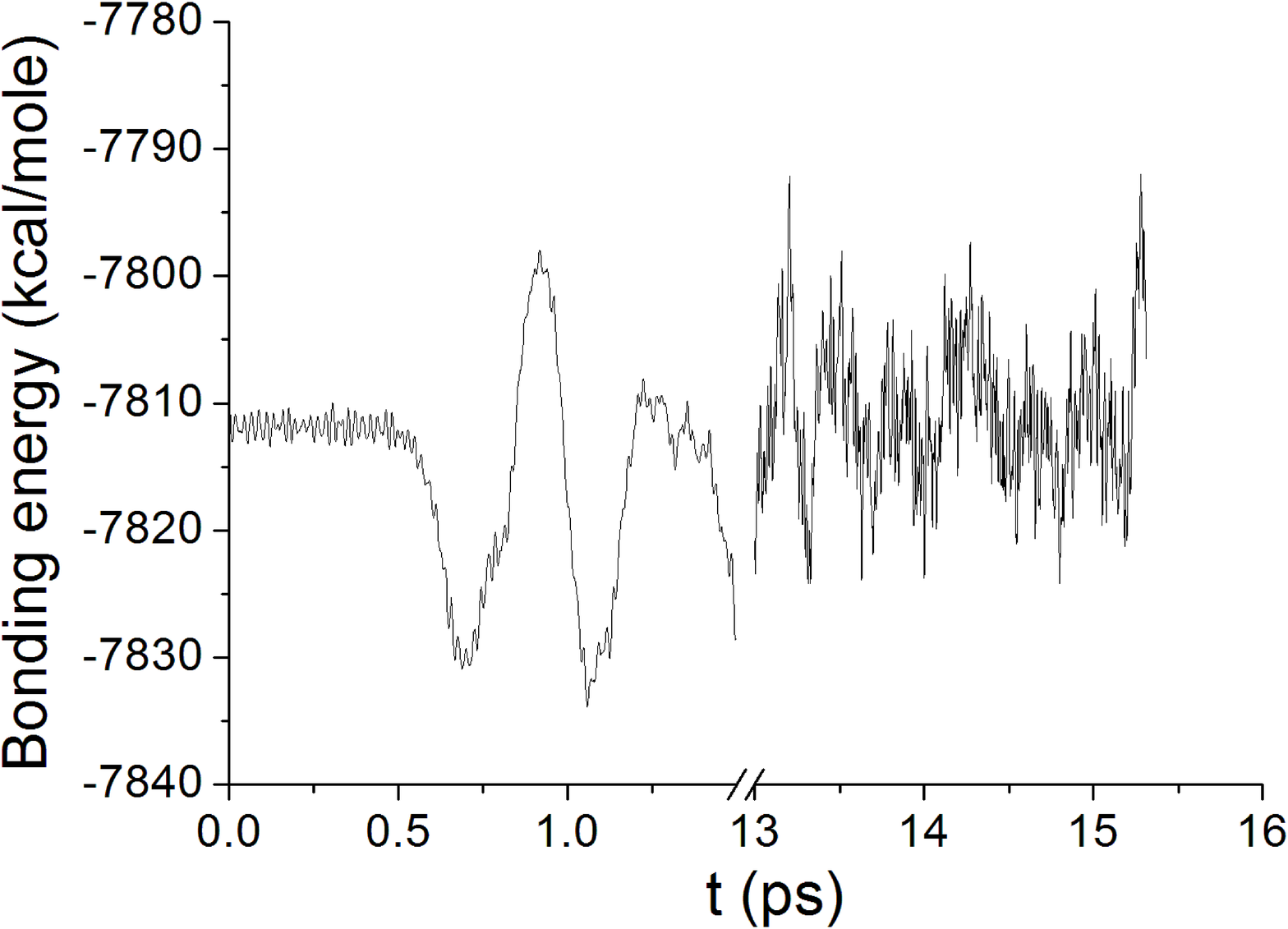}}
\caption[]{The potential energy variations corresponding to Fig. \ref{Fig:DMC12}. In the absence of a magnetic field, the binding energy of the optimized molecule is -7667.1 kcal/mole. Note the break between 2 and 13 ps. During the first period, the curve is regular because  the motion is a nearly pure pendulum oscillation. In the end, the curve is dominated by the much faster internal vibrations, which also induce the random fluctuations of the dipole moment apparent in the same figure.}
\label{Fig:EpotHy}
\end{figure}

Since the lowest phonon frequency of trioS is still much higher than a reasonably high rotation frequency, one is led to use the maximum possible magnetic intensity so as to increase the coupling between them. The molecule is initially at rest in the xz plane and centered at the origin with its principal axis 1 coincident with the Ox coordinate axis. A magnetic field of 50 a.u. is then applied along the Oy axis. This is an unstable configuration and a magnetic torque acts to rotate the molecule about the Ox axis, starting an oscillatory motion. The first period of this (1.5 ps) is represented in Fig. \ref{Fig:DMC12} as the trajectory of the positive tip of the dipole moment (red curve and dots). During this time, the latter remains closely parallel to the yz plane, but already shows signs of tilting. For the sake of clarity, we skip the next 12 ps and show only the last 1.5 ps. This differs in two important respects: a) the tip now describes a curve around the Oy axis, i.e. the field direction, which is a stable orientation for the molecule, if it could be reached; b) the trajectory is obviously much more erratic, a first indication that the dipole moment no longer exactly coincides with the symmetry axis of the molecule due to the increased amplitude of the internal atomic vibrations, which suggests a conversion of rotation into vibrational energy.

\begin{figure}
\resizebox{\hsize}{!}{\includegraphics{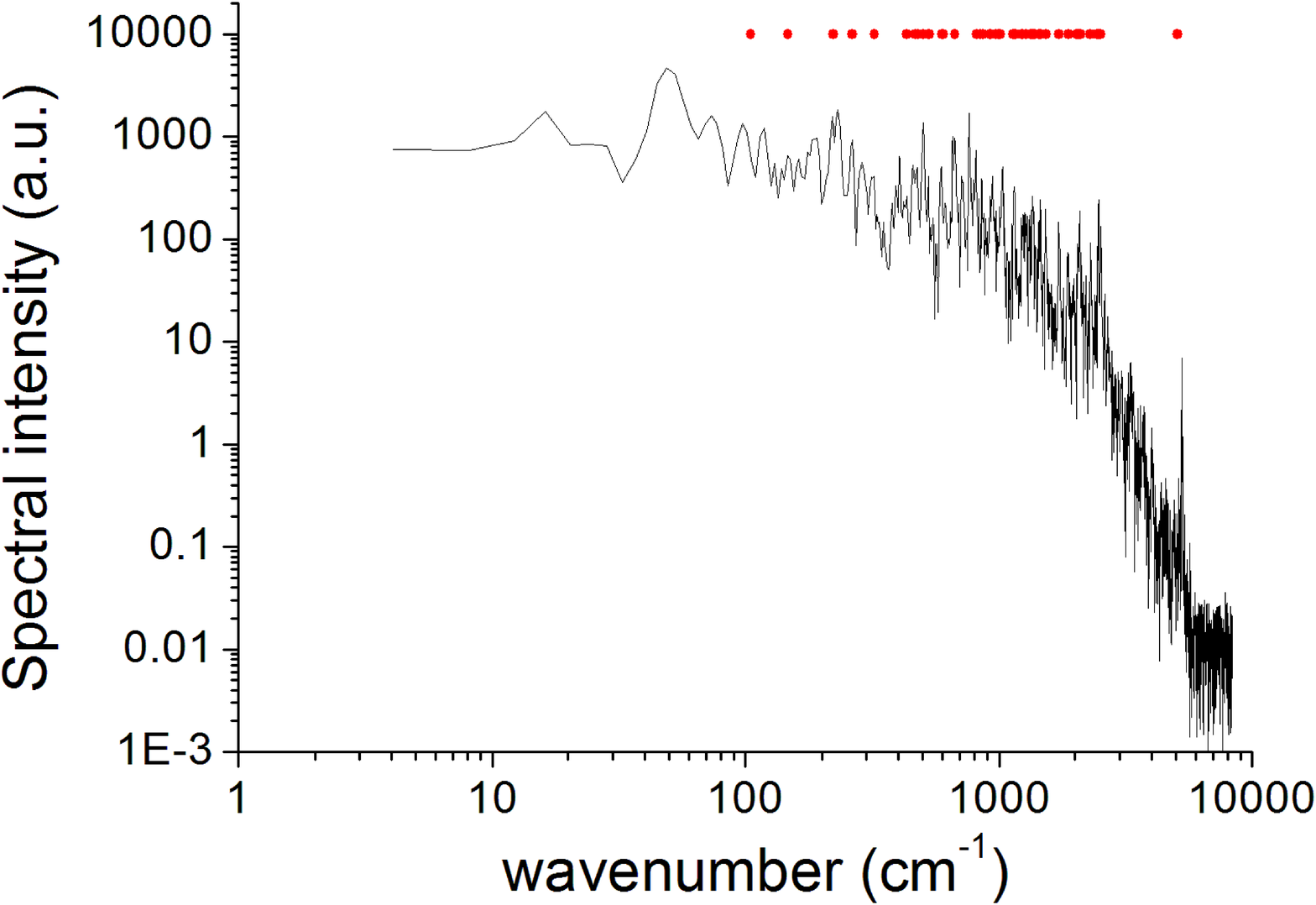}}
\caption[]{FT of the potential energy variations during the last 5 ps of Fig. \ref{Fig:EpotHy} (black line). The red dots at ordinate 10000 represent the IR absorption lines of the molecule. Clearly, the whole vibrational spectrum of the molecule has been excited. The peak near 50 cm$^{-1}$ corresponds to the pseudo-period of the trajectory around the bottom of the stable potential well. Except below 100 cm$^{-1}$ the spectrum is quite close to that of the same molecule heated up to 200 K and left to vibrate in vacuum (no magnetic field). Its excess energy content (kinetic and potential) is then 24 kcal/mole, or about 1eV  This is the clearest evidence of conversion of rotation to internal energy. By comparison, the corresponding spectrum of the same molecule at rest with an internal kinetic temperature of 20 K is much less rich in modes and its peaks do not exceed 100 a.u.}
\label{Fig:specEpHy}
\end{figure}

Another expression of this conversion is apparent in Fig. \ref{Fig:EpotHy} , which displays the variations of the molecular bonding energy during the first and last 2 ps of the experiment (cf. Fig.\ref{Fig:startEpEk}). During the first 0.5 ps, the molecule ``hesitates'' to leave its unstable position, after which it executes one and a half period of relatively regular pendular motion, with exchanges between potential and kinetic energy. Afterwards, this regular oscillation is more and more drowned into the much faster and erratic internal vibrational motions. Since this energy cannot find an oulet, the motion takes the form of a random walk around the bottom of the stable potential well (0, -2.3, 0), the farther from it, the higher the residual kinetic energy.

A more quantitative assessment of the accumulated vibrational energy is provided by a Fourier analysis of the total potential or kinetic energy, as in Fig. \ref{Fig:specEpHy}. Clearly, the whole vibrational spectrum of the molecule (red dots) has been excited. The peak near 50 cm$^{-1}$ corresponds to the pseudo-period of the trajectory around the bottom of the stable potential well. Except below 100 cm$^{-1}$ the spectrum is quite similar to that of the same molecule heated up to 200 K and left to vibrate in vacuum (no magnetic field). Its excess energy content (kinetic and potential) is then 24 kcal/mole, or about 1eV. This is the clearest evidence of conversion of rotation to internal energy.

According to eq. (5), if this experiment were repeated in a field of $5\,10^{-6}$ G, it would last about 3000 s.

\section{Conclusion}
The present study differs from others in three respects: a) it is restricted to the consideration of a single rotation energy dissipation mechanism; b) it applies to both para- and dia-magnetic grains; c) it is illustrated by semi-empirical chemical modeling experiments. 

The said mechanism can be mediated  by Lorentz magnetic or  electric forces, acting on partial electric charges carried by the individual constitutive atoms of the grains.  It converts rotation energy into internal (vibrational) energy: it is an ``internal dissipation" mechanism in the terms of Purcell \cite{pur79}.\rm  The conversion is efficient only if the rotation vector is not parallel to a symmetry axis of the grain and to the field, and provided the grain is large enough and/or the angular frequency high enough that the latter falls within the spectral range of the phonons. The smallest grains cannot meet the latter condition in the ISM.

The Lorentz mechanism differs from anelasticity and IVRET in that it does not consider the grain as an electrically uniformly neutral body (even if it is, as a whole, electrically neutral). The computation of the electronic wave function gives the distribution of electron density over the molecule. The Lorentz force can then be computed on each atom. In this way, the modeling accounts for both anelasticity and electromagnetic forces. As a result, it is observed that, during rotation, energy is  continuously fed into the whole spectrum of internal vibrations, which is not the case with anelasticity alone. At the same time, our experiments indeed exhibit the molecule deformations  (phonon modes) due to the purely mechanical centrifugal forces invoked by Purcell \cite{pur79}. It must be stressed that the vibrational energy is provided by the rotation energy, not the magnetic (or electric) field, although, at the same time, the field also periodically exchanges energy with the molecule through its couple with the magnetic (or electric) moment.

The dissipation time is found to be proportional to the moment of inertia about the relevant principal axis and inversely proportional to the magnetic field, the initial  angular velocity and the number of atoms in the molecule. For an aromatic molecule of 59 carbon atoms, the computation gives $10^{5}$ s in a field of $5\,10^{-6}$  G, which shows that the Lorentz mechanism can contribute significantly to rotation braking. Using this data, it is possible to extrapolate these results to other molecules having similar structures.

With the Lorentz mechanism, it is not possible to distinguish external from internal alignment for both rely on the field and occur simultaneously, contrary to the anelasticity or Barnett mechanisms, for instance (Purcell 1979).

The mechanism is active for both full rotations and pendulum oscillations. It does not depend on the magnetic susceptibility, except that the latter determines the ultimate orientation of the molecule when the rotation energy has been removed. 

Modeling allows the determination of the magnetic susceptibilities along the principal axes. Only one of these corresponds to a stable static configuration when it is parallel to the field; it has one of the two highest moments of inertia. Modeling of the motion in a field shows that, with enough braking, the grain orients itself so it rotates or oscillates about this axis.

Finally, modeling unveils two types of drifts of the grain mass center, also due to the Lorentz force: one accelerated and one with constant speed, depending on the directions of the field and rotation vectors..

As the modeling package provides capabilities for Langevin dynamics at constant temperature, it is possible to study the grain motion in more complex situations, e.g. where the photon influx and/or the ambient H atoms friction cannot be ignored. But this cannot be developed here.

As recalled above, the frequency of the lowest-energy phonon  of the smallest molecules is too high to couple with the rotation frequency, so these molecules cannot align in the field and contribute to polarization. However, this should not be a major handicap for the proposed mechanism, nor, for that matter, for other candidate mechanisms, such as spin-lattice relaxation (Lazarian  and Draine 2000, Hoang et al. 2013). For, large molecules are certainly a major component of dust, as witnessed by the extension of observed dust emission down to a few GHz (see Planck Collaboration 2011-2014 and analysis by Papoular 2015). Also, many PAH models now include quite large PAHs (van Dishoeck 2000, Chiar and Pendleton 2000), and I have shown that the Unidentified Infrared Bands can be emitted by large molecules, PAH and kerogen types (Papoular 2012). Nonetheless, these qualitative considerations need to be backed by an exhaustive quantitative study of the polarizing efficiency of grains as a function of size and shape, a task that remains to be addressed.

\section{Appendix: Drifts in a magnetic field}
 As shown above, the braking mechanism relies upon the coupling between rotation and vibration motions provided by the Lorentz magnetic force. Another manifestation of this force is the drift of the molecule as a whole. There are two kinds of drifts of rotating molecules, according to whether the rotation axis is perpendicular or parallel to molecular axis 3 (orthogonal to the molecular plane). Generic examples are given here.

\subsection{Accelerated drift parallel to the rotation axis}
\begin{figure}
\resizebox{\hsize}{!}{\includegraphics{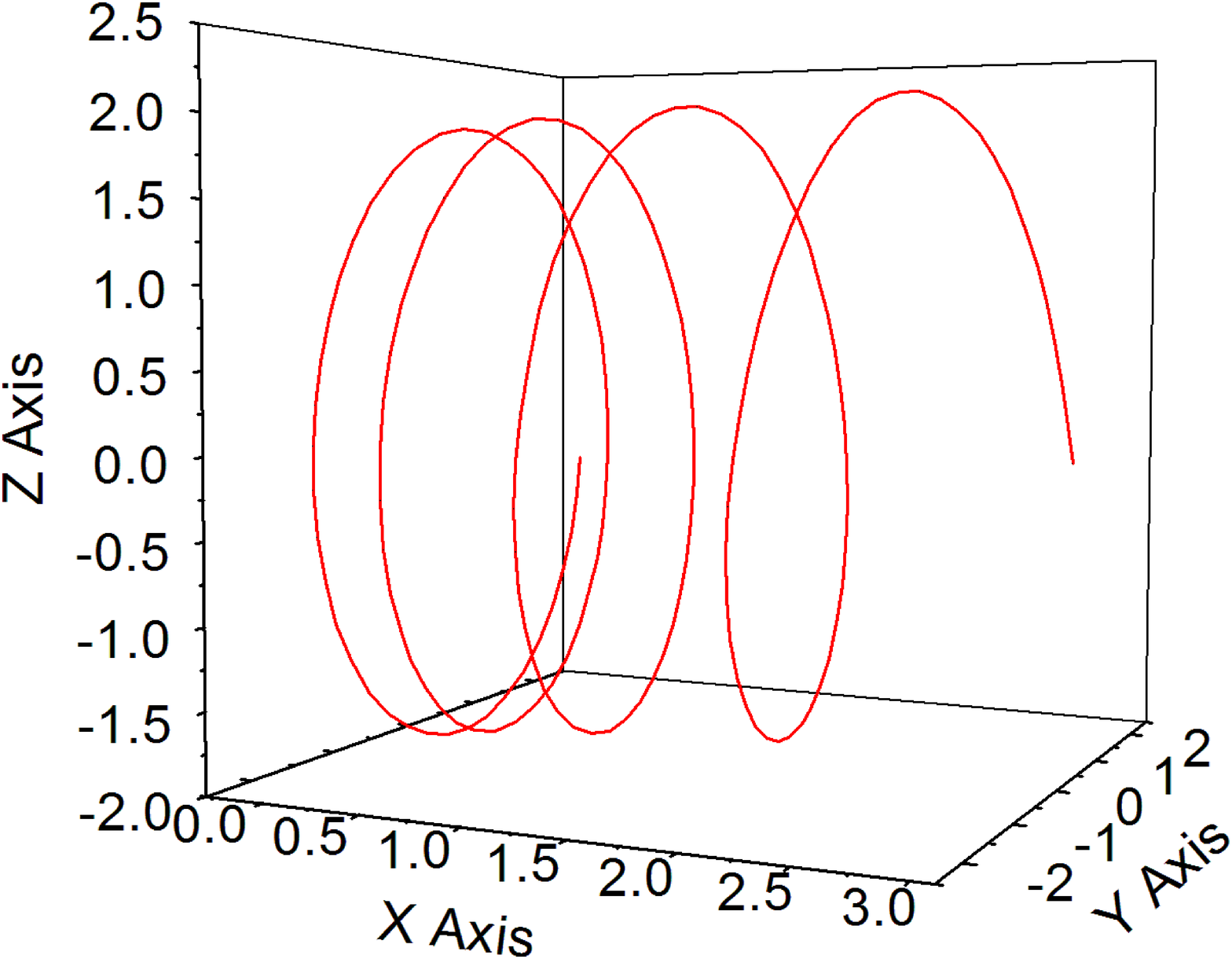}}
\caption[]{Motion of the sulfur atom (tip of the pentagon) in the trioS molecule rotating about its longitudinal principal axis 1 (parallel to Ox), in a field parallel to Oz, with intensity 10 a.u. The initial angular frequency is 4.63 rad/ps and the initial coordinates of S are (0, 1.95, -0.31). The resulting average mass center acceleration over the 5 ps of the experiment is $\sim0.2\,\AA{\ }$/ps$^{2}$, parallel to +Ox (in the same direction as the Lorents force). The increasing kinetic energy is provided by the magnetic field.}
\label{Fig:acceldrift}
\end{figure}

Consider a trioS rotating about its principal axis 1 (coincident with Ox), in a magnetic field parallel to Oz, $H=10$ a.u. The atoms experience a Lorentz force parallel to +Ox or -Ox according to the sign of their net electric charge (higher or lower local electron density). If the molecule has a permanent dipole moment, then the centers of mass of the two sets do not coincide so the resultant force is finite and proportional to both the field and the rotation velocity. Because of the torque between the field and the induced magnetic moment, the rotation velocity periodically increases or decreases according to whether the molecule approaches, or recedes from, the stable magnetic configuration ($H$ parallel to the axis of symmetry). Because the Lorentz force scales like $V$, the acceleration is stronger than the deceleration and, on average, the molecule is uniformly accelerated parallel to the field, as shown in Fig. \ref{Fig:acceldrift}. While, in the braking process, the coupling between rotation and vibrations costs no energy to the magnetic field, the kinetic energy attendant to the drift is provided by the Lorentz force as it is, here, parallel to the drift velocity.

\subsection{Uniform drift orthogonal to the rotation axis}

\begin{figure}
\resizebox{\hsize}{!}{\includegraphics{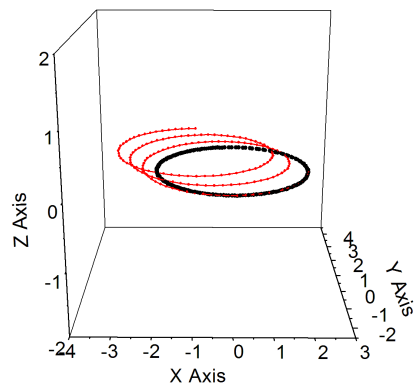}}
\caption[]{Motion of the sulfur atom (tip of the pentagon) in the trioS molecule rotating about its principal axis 3 (parallel to Oz), in a field parallel to Ox, with intensity 20 a.u. The initial angular frequency is 3.47 rad/ps and the initial coordinates of S are (1.97, -0.17, 0 ). The molecule is initially in the xy plane and remains in it during the 5 ps of the experiment, but drifts at constant average velocity, $\sim0.4\,\AA{\ }$ per ps, in both directions -Ox and Oy. The heavy black line is the trajectory of S in the absence of any field. }
\label{Fig:ctdrift}
\end{figure}

Consider a trioS rotating about its principal axis 3 (coincident with Oz), in a magnetic field parallel to Ox, $H=20$ a.u.  In this configuration, the molecule rotates in its own plane and drifts essentially in the same. The molecule drifts with constant (on average) velocities in both directions -Ox and Oy, orthogonally to the rotation vector, as shown in Fig. \ref{Fig:ctdrift}. The cause of this drift is not clear. It may be due to a second-order effect, as it disappears for weak magnetic fields.

Both types of drifts are observed in the experiment of Fig.\ref{Fig:oscil}.

\end{document}